\documentclass[aps,amssymb,amsmath,prl,reprint,noshowpacs]{revtex4-1}
\usepackage{times}
\usepackage{amssymb,amsmath,graphicx}
\usepackage[usenames]{color}
\usepackage{bbold}
\usepackage{siunitx}
\usepackage{braket}

\usepackage[T1]{fontenc}
\usepackage[utf8]{inputenc}
\usepackage[english]{babel}

\newcommand{\imu}{\text{\rm i}}
\newcommand{\diff}[1]{\text{d}}
\usepackage{textcomp}

\newcommand{\figwidth}{0.95\columnwidth} 
\newcommand{\commentOut}[1]{}
\usepackage{scalefnt}   

\newcommand{\affil}{Photonics Laboratory, ETH Zürich, CH-8093 Zürich, Switzerland}

\begin{document}
\scalefont{1.05}
\title{Cold Damping of an Optically Levitated Nanoparticle to micro-Kelvin Temperatures}
\author{Felix Tebbenjohanns}
\affiliation{\affil}
\author{Martin Frimmer}
\affiliation{\affil}
\homepage{http://www.photonics.ethz.ch}
\author{Andrei Militaru}
\affiliation{\affil}
\author{Vijay Jain}
\affiliation{\affil}
\author{Lukas Novotny}
\affiliation{\affil}


\begin{abstract}
We implement a cold damping scheme to cool one mode of the center-of-mass motion of an optically levitated nanoparticle in ultrahigh vacuum (\SI{e-8}{mbar}) from room temperature to a record-low temperature of \SI{100}{\micro K}. 
The measured temperature dependence on feedback gain and thermal decoherence rate is in excellent agreement with a parameter-free model. 
We determine the imprecision-backaction product for our system and provide a roadmap towards ground-state cooling of optically levitated nanoparticles.
\end{abstract}
\date\today

\maketitle

\paragraph{Introduction.}
The interaction of light and matter is at the heart of a host of precision measurements, ranging from the detection of gravitational waves to the definition of the international unit system~\cite{Abbott2016,Mohr2016}. What makes electromagnetic fields our probe of choice is the availability of detectors and laser light sources that operate at the noise limits dictated by the laws of quantum mechanics. Shortly after the invention of the laser, the scientific community started to explore the possibilities of mechanical manipulation of matter using the forces of light in optical traps~\cite{Ashkin1980,Ashkin2006}. 
These forces can be interpreted as the inevitable consequence of the measurement process resulting from light-matter interaction~\cite{Braginsky1992}. Thus, optical forces and measurement precision are linked according to the Heisenberg uncertainty principle.
The investigation of these measurement backaction effects has generated the field of optomechanics, which has developed experimental platforms that allow both measurement and control of mechanical motion at the quantum limit using light fields~\cite{Teufel2009,Anetsberger2010,Verlot2010,Purdy2013,Aspelmeyer2014}.

Dielectric particles levitated in optical traps are particularly versatile optomechanical systems~\cite{Chang2010,Romero-Isart2011a,Li2011,Gieseler2012,Yin2013a}. Next to applications for precision measurements~\cite{Geraci2010,Arvanitaki2013,Moore2014,Rider2016}, one exciting prospect is the investigation and control of quantum states of massive objects~\cite{Romero-Isart2011}. The starting point for any experiment in this direction is cooling an optically levitated nanoparticle to its quantum ground state of motion, a feat already achieved for cryogenically precooled mechanically clamped systems using autonomous cavity cooling~\cite{Chan2011,Teufel2011} and active feedback control~\cite{Rossi2018}. While cavity-based feedback methods have made remarkable progress in recent years~\cite{Kiesel2013,Millen2015,Fonseca2016,Windey2018}, the most successful method to cool the center-of-mass motion of a levitated particle to date has been parametric feedback cooling in a single-beam optical dipole trap~\cite{Gieseler2012,Vovrosh2017}, where cooling from room temperature to occupation numbers below a hundred phonons has been achieved~\cite{Jain2016}. 

Recently, the finite net charge carried by levitated nanoparticles has moved to the center of attention in the context of force sensing ~\cite{Frimmer2017a,Ranjit2015}. Importantly, the Coulomb force that can be applied to a charged optically levitated particle provides the possibility to implement a cooling method termed \emph{cold damping}~\cite{Steixner2005,Bushev2006,Iwasaki2018}. This measurement-based feedback technique applies a direct force to the oscillator in proportion to its speed, effectively leading to an increased damping rate~\cite{Mancini1998}. Cold damping has successfully been used in optomechanics to cool clamped mechanical oscillators~\cite{Cohadon1999,Poggio2007,Wilson2015,Rossi2018} and optically levitated micron-sized particles~\cite{Ashkin1977,Li2011}, using the radiation pressure force. Surprisingly, the potential of cold damping for ground-state cooling the motion of an optically levitated nanoparticle has remained unexplored to date.

In this Letter, we cool the center-of-mass motion of an optically levitated nanoparticle to a temperature of \SI{100}{\micro K} using cold damping. To this end, we exploit the Coulomb force acting on the net electric charge carried by the particle. We investigate the cooling performance as a function of gas pressure and feedback gain to explore the limitations of the method. 
Our system operates a factor of one thousand from the Heisenberg limit of the imprecision-backaction product and provides a platform for studying ground-state cooling of optically levitated oscillators.

\paragraph{Experimental.}
\begin{figure}[htb]
\includegraphics[width=\figwidth]{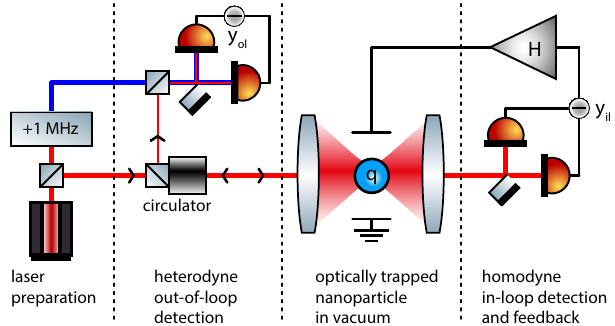}
\caption{Experimental setup. A silica nanoparticle (nominal diameter \SI{136}{nm}) carrying a finite net charge $q$ is optically trapped in vacuum using a laser beam (wavelength \SI{1064}{nm}) focused by an objective. 
To measure the $y$ motion of the particle, the backscattered light is rerouted by a free-space circulator and mixed with a local oscillator (frequency shifted by \SI{1}{MHz} relative to trap laser) to a balanced split detection scheme, yielding the out-of-loop signal $y_\text{ol}$.
The forward scattered light is detected in another balanced split detection scheme and yields the in-loop signal $y_\text{il}$, which is processed by a linear, digital filter $H$. The resulting feedback signal is applied as a voltage to a capacitor enclosing the trapped particle.}
\label{fig:setup}
\end{figure}
Our experimental setup is shown in Fig.~\ref{fig:setup}. We optically trap a silica nanoparticle (diameter 136~nm) in a linearly polarized laser beam (wavelength 1064~nm, focal power 130~mW), focused by a microscope objective (0.85NA) resulting in oscillation frequencies of the particle's center-of-mass $\Omega_{z} = 2\pi\times45~{\rm kHz}$, $\Omega_{x} = 2\pi\times125~{\rm kHz}$, and $\Omega_{y} = 2\pi\times146~{\rm kHz}$, where $z$ denotes the direction along the optical axis, while $x$ ($y$) are the coordinates in the focal plane along (orthogonal to) the axis of polarization. We collect the forward scattered light with a lens and guide it to a standard homodyne detection system for the particle's motion along all three axes, which we call the in-loop detector (only shown for the $y$ axis in Fig.~\ref{fig:setup})~\cite{Gieseler2012}. 
Throughout our work, the particle's motion along the $x$ and $z$ directions is cooled using parametric feedback to temperatures below \SI{1}{K}, rendering non-linearities of the trapping potential irrelevant~\cite{Gieseler2014a,Jain2016}. From here on, we solely focus on the motion of the particle along the $y$ axis. 
We exert a Coulomb force on the net charge carried by the optically trapped nanoparticle by applying a voltage to a pair of electrodes enclosing the trap~\cite{Frimmer2017a}. To cool the particle's motion, this voltage is a feedback signal derived from the measurement signal $y_\text{il}$ acquired from the forward scattered light. 
Our linear feedback filter with transfer function $H(\Omega)$ consists of a series of digital, second-order biquad filters, which essentially mimics a derivative filter, such that the feedback signal is proportional to the particle's velocity. More specifically, we use a band-pass filter whose center-frequency is set to above the particle's oscillation frequency $\Omega_y$, such that the transfer function at $\Omega_y$ increases linearly with frequency while preserving a flat phase response~\cite{Rossi2018}. 
Finally, we measure the out-of-loop signal $y_\text{ol}$ with a heterodyne detection system for the backscattered light, using a local oscillator which is frequency shifted by \SI{1}{MHz} from the trapping light. 
We calibrate our detectors in the mildly underdamped regime at a pressure of \SI{10}{mbar} using the equipartition theorem in the absence of feedback cooling~\cite{Hebestreit2018}.

\paragraph{Cooling performance.}
\begin{figure*}
\includegraphics[width=\textwidth]{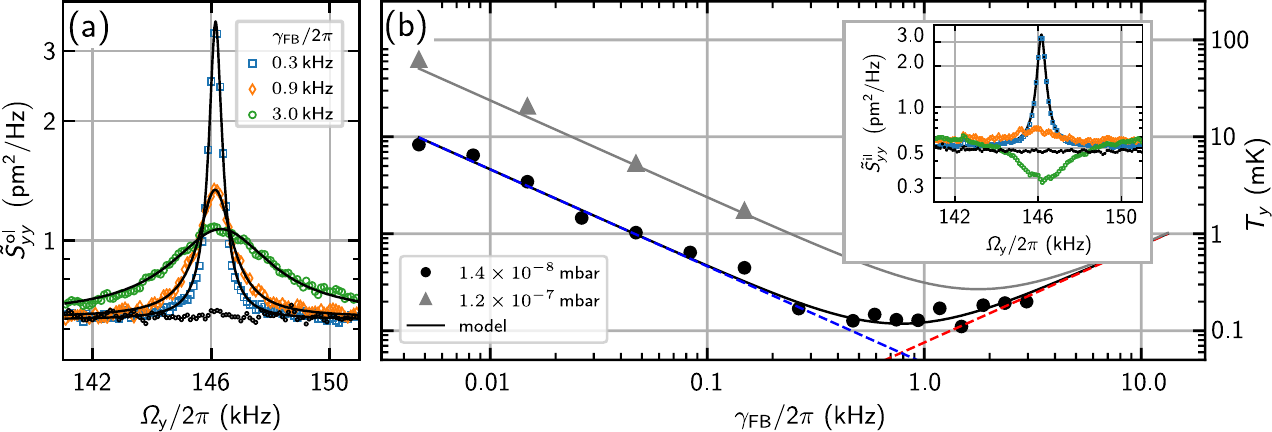}
\caption{
(a)~Single-sided power spectral densities $\tilde{S}_{yy}^\text{ol}$ of the motion of the nanoparticle measured by the out-of-loop detector for different feedback damping rates $\gamma_\text{FB}$.
The solid lines are Lorentzian fits to the data. The black datapoints denote the measured shot-noise level $\tilde{S}_{\nu\nu}$ on the out-of-loop detector.
(b)~Mode temperature $T_y$ derived from the out-of-loop signal $y_\text{ol}$ as a function of feedback gain $\gamma_\text{FB}$. The black circles denote the measured values at a pressure of \SI{1.4e-8}{mbar}. At a damping rate of $\gamma_\text{FB}=2\pi\times1$~kHz, we observe a minimum temperature of \SI{100}{\micro K}. The solid black line is a parameter-free calculation according to Eq.~\eqref{eq:y_power}. The blue (red) dashed line denotes the contribution of the first (second) term in Eq.~\eqref{eq:y_power}. The grey triangles and line show measured and calculated mode temperatures at a higher pressure of \SI{1.2e-7}{mbar}.
Inset: Power spectral densities $\tilde{S}_{yy}^\text{il}$ measured by the in-loop detector for the same settings as in (a). Photon shot-noise $\tilde{S}_{nn}$ is shown as black datapoints. In contrast to (a), for large feedback gain ($\gamma_\text{FB}=2\pi\times3.0$~kHz) we observe noise squashing, \emph{i.e.}, the measured signal drops below the noise floor.
}

\label{fig:e_vs_gain}
\end{figure*}
We now investigate the performance of our cold damping scheme at a pressure of \SI{1.4e-8}{mbar}. In Fig.~\ref{fig:e_vs_gain}(a) we show the single-sided power spectral density (PSD) $\tilde{S}_{yy}^\text{ol}$
\footnote{We define our PSDs according to $\Braket{y^2} = \int_0^{\infty} {\rm d}f ~\tilde{S}_{yy}(f) = \int_{-\infty}^{\infty} {\rm d}\Omega ~ S_{yy}(\Omega)$}
of the out-of-loop signal for different feedback gains, which we express as damping rates $\gamma_\text{FB}$.
We extract the damping rate $\gamma_\text{FB}$ from ring-down measurements as detailed further below.
The measured signal $\tilde{S}_{yy}^\text{ol}$ corresponds to a Lorentzian function added to a spectrally flat noise floor due to the photon shot noise on our detector. The spectral width of the Lorentzian is a measure for the total damping rate arising from feedback cooling and residual gas damping. The latter is largely negligible under feedback at the low gas pressures of our experiments. The area under the Lorentzian, on the other hand, is a measure for the energy (\emph{i.e.}, temperature) of the particle's oscillation mode. As expected, as we increase the feedback gain, the Lorentzian broadens in width and simultaneously shrinks in area. 
Thus, from the PSD of the out-of-loop signal $\tilde{S}_{yy}^\text{ol}$, we extract the energy $k_B T_y$ in the $y$ mode of the levitated particle.

In Fig.~\ref{fig:e_vs_gain}(b), we plot the measured mode temperature $T_y$ as a function of feedback damping rate $\gamma_\text{FB}$ at a pressure of \SI{1.4e-8}{mbar} as black circles. At small feedback gains, we observe a decrease in oscillator temperature with increasing feedback gain. However, there exists an optimal feedback gain of about \SI{1}{kHz}. 
For gain values larger than the optimum, the oscillator temperature increases with increasing feedback gain. For comparison, we show the PSD of the measured in-loop signal $\tilde{S}_{yy}^\text{il}$ in the inset of Fig.~\ref{fig:e_vs_gain}(b) at the same gain values as in Fig.~\ref{fig:e_vs_gain}(a). For large feedback gain, we observe that $\tilde{S}_{yy}^\text{il}$ drops below the shot noise level. This effect, termed \emph{noise squashing}, arises from correlations between the particle's position and the measurement noise that is fed back by the control loop~\cite{Poggio2007,Wilson2015}.
In Fig.~\ref{fig:e_vs_gain}(b), we additionally show measurements performed at a higher pressure of \SI{1.2e-7}{mbar}
(grey triangles), where the increased gas damping rate leads to a larger mode temperature as compared to the low-pressure data.

\paragraph{Analysis.}
To understand our results, let us analyze our system from a theoretical perspective. The Fourier transform $\hat y(\Omega)$ of the time-dependent particle position $y(t)$ follows the equation of motion
\begin{equation}
\label{eq:eq_of_motion_ft_closed}
\hat{y} \left[\Omega_y^2 - \Omega^2 + \imu \gamma \Omega - H(\Omega)\right] = \frac{\hat{f}_\text{fluct}}{m} + H(\Omega) \hat{y}_n ,
\end{equation}
where $\Omega_y$ is the $y$ mode's eigenfrequency, $m$ the particle's mass, and $\hat{y}_n$ is the measurement shot noise on the in-loop detector, which measures $\hat y_\text{il}=\hat y+\hat y_n$. The damping rate $\gamma$ arises from the interaction with residual gas molecules. The term $\hat f_\text{fluct}$ describes the fluctuating force generated by the interaction with the gas and from radiation pressure shot noise. Via the fluctuation dissipation theorem, $\hat f_\text{fluct}$ is inextricably linked to $\gamma$~\cite{Clerk2010}. Within the bandwidth of interest, the transfer function of our feedback circuit is well described by $H(\Omega) = -\imu\gamma_\text{FB}\Omega$. The feedback damping rate $\gamma_\text{FB}$ can be set by adjusting the feedback gain and incorporates the exact geometry of the capacitor electrodes and the number of charges carried by the levitated particle. Importantly, the feedback transfer function $H(\Omega)$ appears twice in Eq.~\eqref{eq:eq_of_motion_ft_closed}, which results from the fact that the input to the feedback circuit $\hat{y}_\text{il}$ is the sum of the true position $\hat y$ and the measurement shot noise $\hat{y}_n$.
From Eq.~\eqref{eq:eq_of_motion_ft_closed}, we obtain the two-sided PSD 
on the out-of-loop detector

\begin{equation}\label{eq:S_yy}
S^\text{ol}_{yy}(\Omega) = \frac{S_{ff}/m^2 + \gamma_\text{FB}^2 \Omega^2 S_{nn} }{\left(\Omega_y^2-\Omega^2\right)^2 + \left(\gamma+\gamma_\text{FB}\right)^2 \Omega^2} + S_{\nu\nu},
\end{equation}
where $S_{ff}$ denotes the PSD of the fluctuating force $\hat f_\text{fluct}$, and $S_{nn}$ ($S_{\nu\nu}$) are the PSDs of the in-loop (out-of-loop) detector noise.
Integrating the first term of Eq.~\eqref{eq:S_yy}, which corresponds to the PSD of the true position $y$, in the limit $\gamma_\text{FB}\gg\gamma$ yields the variance 
\begin{equation}\label{eq:y_power}
\Braket{y^2} = \frac{\pi S_{ff}}{m^2 \gamma_{FB} \Omega_y^2} + \pi \gamma_{FB} S_{nn},
\end{equation}
which is a direct measure for the temperature $T_y=m\Omega_y^2\Braket{y^2}/k_B$ of the oscillator mode. The first term contributing to the expression in Eq.~\eqref{eq:y_power} scales with the inverse of the feedback cooling rate $\gamma_\text{FB}$. This term resembles the desired action of the feedback, which is to reduce the impact of the heating term given by the fluctuating force $S_{ff}$. Importantly, the second term is proportional to the feedback damping rate, which multiplies with the measurement noise $S_{nn}$. This term resembles the undesired but inevitable effect of the control loop heating the particle by feeding back measurement noise.
Accordingly, our model predicts the existence of an optimum feedback cooling rate, where the mode temperature reaches its minimum value $T_\text{min} = 2\pi\Omega_y\sqrt{S_{ff}S_{nn}}/k_B$, a behavior that we observe in our measurements in Fig.~\ref{fig:e_vs_gain}(b). 

\begin{figure}
\includegraphics[width=\figwidth]{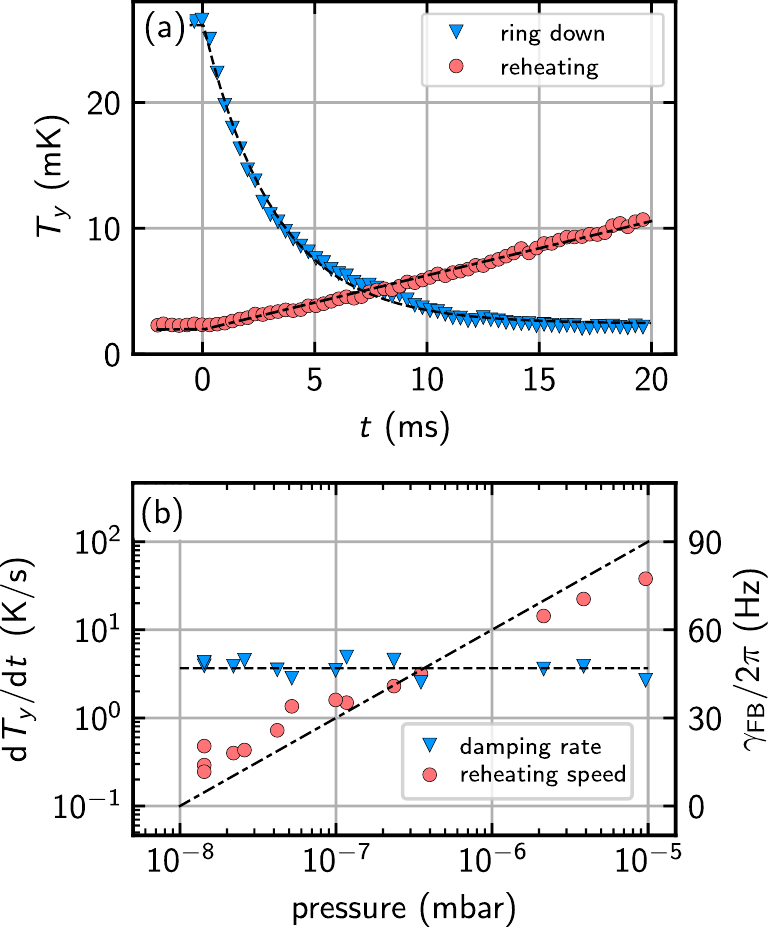}
\caption{
(a)~Ring-down and reheating experiment. For the ring-down experiment, we start with the oscillation mode at an elevated temperature, reached by reducing the feedback gain. At time $t=0$, we switch the feedback damping rate to $\gamma_\text{FB}$ and measure the decay of the mode temperature $T_y(t)$ (blue triangles). We fit $T_y$ with a single exponential decay (black dashed line) and extract the decay constant, which yields $\gamma_\text{FB}=2\pi\times47~{\rm Hz}$. 
For the reheating experiment, we turn off the feedback-cooling at time $t=0$ and measure the increasing mode temperature $T_y(t)$ (red circles). A linear fit (dash-dotted line) to the data yields the reheating speed ${\rm d}T_y/{\rm d}t=\gamma T_\text{bath}$.
(b)~Feedback damping rate $\gamma_\text{FB}$ (blue triangles) and reheating speed ${\rm d}T_y/{\rm d}t$ (red circles) as a function of pressure. The feedback damping rate is independent of pressure and solely determined by the gain of the feedback circuit. Within our pressure range the reheating follows a linear trend (indicated as the dash-dotted line).
}
\label{fig:ring_up_down}
\end{figure}

For a quantitative comparison of measurement and theory, we have to determine all parameters entering Eq.~\eqref{eq:y_power}. We extract the in-loop measurement noise $S_{nn}$ from the PSD shown in the inset of Fig.~\ref{fig:e_vs_gain}(b). To obtain the feedback damping rate $\gamma_\text{FB}$, we perform ring-down measurements. To this end, we toggle the feedback gain back and forth between $\gamma_\text{FB}$ for \SI{30}{\micro s} and a much lower feedback gain $\gamma_\text{FB}^\text{low} = \gamma_\text{FB}/300$ for \SI{50}{\micro s}. As shown in Fig.~\ref{fig:ring_up_down}, we measure the mode temperature as a function of time after the gain was switched from $\gamma_\text{FB}^\text{low}$ to $\gamma_\text{FB}$ at time $t=0$. The blue triangles in Fig.~\ref{fig:ring_up_down}(a) are the ensemble average over 100 such decay curves. We observe an exponential decay of the temperature and extract its time constant, which equals $\gamma_\text{FB}$. 
When the feedback gain is switched from $\gamma_\text{FB}$ to $\gamma_\text{FB}^\text{low}$ at time $t=0$, we observe the mode temperature increasing linearly in time [red circles in Fig~\ref{fig:ring_up_down}(a), averaged over 100 reheating experiments]. Since the observed time is much shorter than the inverse damping rate $\gamma$, we expect the temperature to increase as $T(t)=\gamma T_\text{bath}~t$. Together with the fluctuation dissipation theorem $S_{ff} = m\gamma k_B T_\text{bath}/\pi$, the measured slope of the reheating curve therefore provides us with a direct measurement of the first term in Eq.~\eqref{eq:y_power}~\cite{Clerk2010}. Equipped with the experimentally determined values for $\gamma_\text{FB}$, $S_{nn}$, and $S_{ff}$, we calculate the mode temperature as a function of feedback gain according to Eq.~\eqref{eq:y_power} and display it as the solid black line in Fig.~\ref{fig:e_vs_gain}(b). The dashed lines show the two separate contributions from the bath (blue) and measurement noise (red) to Eq.~\eqref{eq:y_power}. Our model describes our experimental findings very well. We stress that there is no free parameter or fit involved.

Finally, we investigate the reheating speed and the ring-down rate $\gamma_\text{FB}$ as a function of pressure. The results are displayed in Fig.~\ref{fig:ring_up_down}(b). We find that the ring-down rates (blue triangles) do not depend on pressure. This observation confirms that the damping rate under feedback is indeed fully dominated by and therefore equivalent to the cold-damping rate $\gamma_\text{FB}$. 
The red circles in Fig.~\ref{fig:ring_up_down}(b) show the measured reheating speeds ${\rm d}T_y/{\rm d}t$ as a function of pressure, which follows the expected linear behavior (dash-dotted line). 

\paragraph{Discussion.}
Let us discuss the current limitations and future prospects of our cold-damping approach for levitated optomechanics. 
To this end, we return to Eq.~\eqref{eq:y_power}, whose two contributions are fundamentally related by the imprecision-backaction product $S_{ff} S_{nn} = \left( \frac{\hbar}{4\pi}\right)^2 \frac{1}{\eta}$, with the measurement efficiency $\eta\le 1$ \cite{Clerk2010}. At the optimal feedback gain, we find an effective phonon occupation number $n=k_BT/(\hbar\Omega_y)-1/2$ that solely depends on $\eta$ as $n_\text{min} = \frac{1}{2}(\frac{1}{\sqrt{\eta}} - 1)$.
At the Heisenberg limit of unit efficiency $\eta=1$, when the fluctuating force $S_{ff}$ driving the system under investigation is purely due to measurement backaction, and the imprecision noise $S_{nn}$ is minimized by optimally detecting all photons scattered by the levitated particle, the particle's motion could, in principle, be brought to its quantum ground state $n_\text{min} = 0$.  In our case, at the lowest investigated pressure of $\SI{1.4e-8}{mbar}$, we extract a total efficiency of $\eta=9\times10^{-4}$ and hence an occupation number of about 16. Our measurements in Fig.~\ref{fig:ring_up_down}(b) suggest that we can further reduce $S_{ff}$ by moving to even lower pressures, before entering the regime where reheating is fully dominated by photon recoil~\cite{Jain2016}. 
The factor $S_{nn}$ in our case is limited by the finite collection and detection efficiency. 
The latter is restricted by the non-ideal mode-overlap between the scattered dipole field and the Gaussian trapping beam on the detector.
Exploiting the Purcell-enhanced collection and detection efficiency of a cavity, a suppression of $S_{nn}$ by more than one order of magnitude seems realistic~\cite{Kiesel2013,Fonseca2016,Windey2018}. Accordingly, occupation numbers approaching unity appear within reach.

\paragraph{Conclusion.}
In conclusion, we have demonstrated cold damping of the center-of-mass motion of an optically levitated nanoparticle from room temperature to \SI{100}{\micro K}, corresponding to less than 20 phonons. We have determined the optimal feedback-damping rate for our system, in excellent agreement with a parameter-free model. Together with photonic techniques under development~\cite{Windey2018,Kuhn2017}, our results put ground-state cooling of optically levitated nanoparticles firmly within reach. 
Besides setting a new temperature benchmark, we believe that our feedback control scheme will serve as a model system for the levitated optomechanics community. 
Putting our work into context, our approach is complementary to parametric feedback cooling, the method of choice to control charge-neutral optically levitated particles. In contrast, our system relies on the levitated object carrying finite net charge. 
Importantly, our work provides the direct connection to established optomechanical technologies~\cite{Poggio2007,Wilson2015,Rossi2018}. This fact generates the opportunity to leverage the insights gained with mechanically clamped systems to drive levitated optomechanics forward.

\commentOut{Furthermore, our work suggests a paradigm change for feedback cooling \textcolor{blue}{towards the ground state} in the context of levitated optomechanics. Before our current work, parametric feedback-cooling has been the most powerful cooling method for optically levitated nanoparticles~\cite{Jain2016}. Unfortunately, the lack of a simple but accurate theoretical description required the community to rely largely on heuristic modeling. 
}

\begin{acknowledgments}
This research was supported by ERC-QMES (Grant No. 338763) and the NCCR-QSIT program (Grant No. 51NF40-160591). We thank R.~Diehl, E.~Hebestreit, F.~van der Laan, R.~Reimann, and D.~Windey for valuable input and discussions.
\end{acknowledgments}

%


\end{document}